\documentclass[12pt]{article}
\begin{document}

\begin{center}{\bf\Large The Concept of Dark Energy is not Based on the Principles of Physics: Cosmological Data Can be clearly Explained Without this Concept}\end{center}

\begin{center}Felix M Lev\end{center}
\begin{center} Independent Researcher, San Diego, 92010 CA USA\end{center}
\begin{center}Email: felixlev314@gmail.com\end{center}
\begin{abstract}
The physics community has adopted the principle that when new experimental data appears, physicists should first try to explain it based on existing science.
Only if all such attempts fail can new exotic explanations be brought in. However, in the case of cosmological acceleration, the opposite approach was taken: without serious attempts to explain this phenomenon from existing science, physicists attracted dark energy and other exotic concepts whose physical meaning is a mystery.
As shown in our publications, the cosmological acceleration can be clearly explained without uncertainties. The derivation of this explanation requires significant technical
efforts described in our publications. The purpose of this article published in the Proceedings of the 2nd International Conference on Gravitation, Astrophysics and Cosmology (April 16-17, 2026, Paris, France) is to explain our approach at the simplest possible level so that the basic ideas of our approach would be understandable to many physicists and astrophysicists.
\end{abstract}

\begin{flushleft} Keywords: irreducible representations; cosmological acceleration; de Sitter symmetry \end{flushleft}

\section{
Statement of the problem of cosmological acceleration}
\label{intro}

Let's consider a system of macroscopic bodies that are located at large distances from each other so that all interactions between the bodies (gravitational, electromagnetic and others) can be neglected. Let us also assume that the sizes of these bodies are much smaller than the distances between them. Then, the problem of describing such a system of bodies seems obvious: since all bodies are at large distances from each other, the motion of each body as a whole does not depend on other bodies and each body can only move at some constant speed with zero acceleration. 

However, physicists were surprised when in 1998 observations \cite{Perlmutter} showed that the bodies move relative to each other with the relative acceleration
\begin{equation}
{\bf a}={\bf r}c^2/R^2
\label{Perl}
\end{equation}
where ${\bf r}$ is the relative radius-vector and $R$ is a quantity with the dimension of length.
Usually this quantity is expressed in terms of the cosmological constant $\Lambda$ as $\Lambda =3/R^2$ and the recent observational data of the Planck collaboration \cite{Planck} show that $\Lambda = 1.3\cdot 10^{-52}/m^2$ with the accuracy 5\%. Therefore $R$ is a quantity of the order of $10^{26}m$. 
Thus, observations have shown that bodies repel each other, and the repulsive force is proportional (not inversely proportional) to the distance between them. The formula (\ref{Perl}) also shows that in our daily life and even in the Solar System, this repulsive force is negligible. However, it becomes significant for bodies located at cosmological distances from each other.

The first impression might be that the result (\ref{Perl}) follows from General Relativity (GR) if we assume that we live in de Sitter (dS) space with the radius of curvature $R$ and then, as is well known, $\Lambda$ is treated as the curvature of dS space. However, the following questions arise: a) why nature preferred dS space and not some other space; b) why the value of $R$ is as observed and not some other value; c) why physicists have decided that the real background space should be flat (that is, that in nature there should be $\Lambda=0$). In Sec. \ref{Dark} we describe the history of this problem and explain  why the mainstream literature describes cosmological acceleration in terms of exotic mechanisms (dark energy or quintessence and others). In Sec. \ref{Hierarchy} we explain why the cosmological constant problem
should be treated not from GR or other field theories but from the point of view of quantum theory in semiclassical approximation.

\section{Why physicists decided to describe cosmological data with dark energy}
\label{Dark}

It is well known that Einstein originally thought that our universe was stationary. Then, as follows from his equations, this can only be if $\Lambda\neq 0$
\cite{EinsteinLambda}. However, when Einstein visited Hubble at his Mount Wilson Observatory and Hubble showed him the results of observations \cite{Hubble}, 
then, according to Gamov's recollections, Einstein said that his statement that $\Lambda\neq 0$ was the biggest blunder of his life. 
After that, the mainstream literature (including textbooks) began to claim that $\Lambda=0$ is a necessary condition. The main argument in favor of this statement was that  curvature of space-time is created by matter and so, the empty space-time should be the flat Minkowski space. 

Let us discuss this argument from the point of view of the principles of physics. According to these principles, the definition of any physical quantity is a description of the way of measuring this quantity. From this point of view, the question arises what physical meaning the background space-time has and why many physicists believe that we live in some kind of background space. When there are many particles in the world, it may seem that they are in some kind of space-time background.
However, coordinates of background space are not directly measurable physical quantities, only particle coordinates are such quantities (in the approximation when well-defined operators of such coordinates exist). 

Therefore, the question arises whether the concept of background space itself is fundamental and whether 
fundamental physical theory needs this concept at all. In classical field theories, the coordinates of background space do not have a direct physical meaning since they are not directly measured but indirectly they have a physical meaning: they are built into the mathematical technique that allows to calculate the coordinates of real particles.

From a mathematical point of view, one can also consider the case of empty spaces, that is, spaces in which there are no particles. However, these cases have no physical meaning since it is impossible to carry out measurements in spaces that exist only in our imagination. 
Therefore, discussions about whether empty space can be curved or flat also have no physical meaning.

Let us also make the following remarks. At present, there are no universal theories in physics that work for all values of the parameters included in such theories. 
For example, nonrelativistic theory cannot be extrapolated to cases when speeds are comparable to $c$ and classical physics cannot be extrapolated for describing energy levels of the hydrogen atom. GR is a successful classical (non-quantum) theory for describing macroscopic phenomena where large masses (stars and planets) are present, but the extrapolation of GR to the case of empty space is not physical. 

The claim that $\Lambda$ must be zero has also been criticized by several authors (e.g. the authors of \cite{Rovelli} titled “Why All These Prejudices Against a Constant?”) on the following grounds: 
GR without the contribution of $\Lambda$ has been confirmed with a good accuracy in experiments in the Solar System. If $\Lambda$ is as small as it has been observed then it can have a significant effect only at cosmological distances while for experiments in the Solar
System the role of such a small value is negligible. That is why it is not clear why we
should think that only a special case $\Lambda=0$ is allowed. If we accept the theory containing a gravitational constant G which cannot be calculated and is taken from the outside then why can’t we accept a theory containing two independent constants?

Despite these obvious facts, the question of the curvature of empty space is discussed in the physics literature and, even in textbooks, it is stated that empty space can only be flat. That is why, the fact that the results \cite{Perlmutter,Planck} could be described only
with $\Lambda\neq 0$ was first perceived as a shock of something fundamental. 

However, the following way out of this situation was proposed:
the terms with $\Lambda$ in the Einstein equations have been moved from the l.h.s. to the 
r.h.s. and were interpreted not as the curvature of empty space-time (which was supposed to be zero), but  as a manifestation of hypothetical fields called dark energy or quintessence. 
Although their physical nature remains a mystery (see e.g., \cite{Brax} and references therein), 
and, as noted in \cite{Kam1}, there are an almost endless number of explanations
for dark energy, mainstream publications on the problem of cosmological acceleration (PCA) involve those concepts. However, these approaches have not solved PCA without uncertainties.

While in most publications, only proposals about 
future discovery of dark energy are considered, the authors of \cite{Brax} stated
that dark energy had already been discovered by the XENON1T collaboration.
In June 2020, it reported an excess of electron recoils: 285 events, 53 more than expected 232 with a statistical significance of 
$3.5\sigma$. However, in July 2022, a new analysis by the XENONnT collaboration \cite{arxiv}
discarded the excess.

As noted below, at the current stage of the universe (when semiclassical approximation is valid), PCA can be explained without uncertainties and without involving 
models and/or assumptions containing ambiguities.

\section{Problem of hierarchy of physical theories}
\label{Hierarchy}

From a theoretical point of view, quantum theory is more general than classical theory: the latter is a special case of the former in the formal limit $\hbar\to 0$. 
Therefore, any result of the classical theory can be derived in quantum theory in the semiclassical approximation in the limit $\hbar\to 0$.
Analogously, since relativistic theory is more general than nonrelativistic one, any result of the latter can be derived in the former in the nonrelativistic approximation
$c\to\infty$. However, these facts do not mean that in all cases we should use only more general theories. For example, if we apply relativistic theory to describe everyday life phenomena in which all speeds are much less than $c$, then such a description will turn out to be technically quite complex. Similarly, it is believed that to describe the kinematics of large macroscopic bodies in the universe it is sufficient to use classical theory, since the use of quantum theory would lead to large unjustified complications.

However, as we saw in the preceding section, the application of classical theory to describe PCA faces the following problems:
\begin{itemize}
\item a) Is it necessary to require that $\Lambda$ must be zero?
\item b) If $\Lambda \neq 0$ then GR does not tell us why nature chose this value of $\Lambda$ and not another.
\item c) If, for some reason, $\Lambda\neq 0$ then is $\Lambda$ a fundamental constant that has the same value at all stages of the evolution of the universe?
\end{itemize}

Quantum theory is more preferable than classical theory even from a purely logical point of view: quantum theory describes particles and there is no situation where the background space is empty. We will see below that quantum theory gives clear answers to the questions a)-c).
For example, as already noted, nonrelativistic theory is a special degenerate case
of relativistic theory in the limit $c\to\infty$ and classical theory is a special degenerate case
of quantum theory in the limit $\hbar\to 0$. These cases are discussed in the literature using many examples. However, a question arises: is it possible to give a general criterion when theory {\it A} is more general than theory {\it B}, and theory {\it B} is a special degenerate case of theory {\it A}? In \cite{book} and our other publications, we proposed the following criterion:
 
{\bf Definition: }{\it Let theory A contain a finite nonzero parameter and theory B be obtained from theory A in the formal limit when the parameter goes to zero or infinity. Suppose that with any desired accuracy theory A can reproduce any result of theory B by choosing a value of the parameter. On the contrary, when the limit is already taken, one cannot return back to theory A and theory B cannot reproduce all results of theory A. Then theory A is more general than theory B and theory B is a special degenerate case of theory A}.

In particular, as shown e.g., in \cite{book} this means that:
\begin{itemize}
\item Any result of nonrelativistic theory can be obtained with any desired accuracy from  relativistic theory with some choice of $c$. On the other hand, in nonrelativistic theory it is not possible to obtain those results of relativistic theory where it is crucial that
$c$ if finite and not infinitely large.
\item Any result of classical (non-quantum) theory can be obtained with any desired accuracy from  quantum theory with some choice of $\hbar$. On the other hand, in classical theory it is not possible to obtain those results of quantum theory where it is crucial that
$\hbar$ if finite and not infinitely small.
\end{itemize}

In the literature, symmetry in quantum field theory (QFT) is usually explained as follows.
Since Poincare group is the group of motions of Minkowski space, the system under consideration should be described by unitary representations of this group. 
This implies that the representation generators are selfadjoined and commute according to the commutation relations of the Poincare group Lie algebra:
\begin{eqnarray}
	&[P^{\mu},P^{\nu}]=0,\quad [P^{\mu},M^{\nu\rho}]=-i(\eta^{\mu\rho}P^{\nu}-
	\eta^{\mu\nu}P^{\rho}),\nonumber\\
	&[M^{\mu\nu},M^{\rho\sigma}]=-i (\eta^{\mu\rho}M^{\nu\sigma}+\eta^{\nu\sigma}M^{\mu\rho}-
	\eta^{\mu\sigma}M^{\nu\rho}-\eta^{\nu\rho}M^{\mu\sigma})
	\label{PCR}
\end{eqnarray}
where $\mu,\nu=0,1,2,3$, $\eta^{\mu\nu}=0$ if $\mu\neq \nu$, $\eta^{00}=-\eta^{11}=
-\eta^{22}=-\eta^{33}=1$,
$P^{\mu}$ are the operators of the four-momentum and  $M^{\mu\nu}$ are the operators of Lorentz angular momenta. This approach is in the spirit of 
the Erlangen Program proposed by Felix Klein in 1872 when quantum theory did not yet exist.
However, although the Poincare group is the group of motions of Minkowski space, 
the description (\ref{PCR}) does not involve this group and this space.

As indicated above, background space is only a mathematical concept: in quantum theory, each physical quantity should be described by an operator but there are no operators for the coordinates of background space. {\it There is no law that every physical theory must contain background space.} For example, it is not used in nonrelativistic quantum mechanics and in irreducible representations (IRs) describing elementary particles. In particle theory, transformations from the
Poincare group are not used because, according to the Heisenberg $S$-matrix program, it is possible to describe only transitions of states from the infinite past when $t\to -\infty$ to the distant future 
when $t\to +\infty$.  In this theory, systems are described by observable physical quantities --- momenta and angular momenta. So, {\it symmetry at the quantum level is defined not by a background space and its group of motions but by 
the condition that the commutators of the operators describing the system under consideration are determined by the symmetry algebra of this system}.	In particular, Eqs. (\ref{PCR}) can be treated as the definition of relativistic (Poincare) invariance at the
quantum level. 

Then each elementary particle is described
by a selfadjoint IR of a real Lie algebra $A$ and a system of $N$ noninteracting particles is
described by the tensor product of the corresponding IRs. This implies that, for the system as a whole, each momentum operator  is a sum of the corresponding single-particle momenta, 
each angular momentum operator is a sum of the corresponding single-particle angular momenta, and {\it this is the most complete possible description of this system}. 
In particular, nonrelativistic symmetry implies that $A$ is the Galilei algebra, relativistic (Poincare) symmetry implies that $A$ is the Poincare algebra, de Sitter (dS) symmetry implies that $A$ is the dS algebra so(1,4) and anti-de Sitter (AdS) symmetry implies that $A$ is the AdS algebra so(2,3).

In his famous paper "Missed Opportunities" \cite{Dyson} Dyson notes that: 
\begin{itemize}
	\item a) Relativistic
	quantum theories are more general than nonrelativistic quantum theories
	even from purely mathematical considerations because Poincare group is more symmetric
	than Galilei one: the latter can be obtained from the former by contraction $c\to\infty$. 
	\item b) dS and AdS 
	quantum theories are more general than relativistic quantum theories
	even from purely mathematical considerations because dS and AdS groups are more symmetric
	than Poincare one: the latter can be obtained from the former by contraction $R\to\infty$
	where $R$ is a parameter with the dimension $length$, and the meaning of this parameter
	will be explained below.   
	\item c) At the same time, since dS and AdS groups are semisimple, they have a maximum possible symmetry and cannot be obtained from more symmetric groups by contraction. 
\end{itemize}

As noted above, symmetry at the quantum level should be defined by a symmetry algebra for the system under consideration. In \cite{book}, the
statements a)-c) have been reformulated in terms of the corresponding Lie algebras and it has
also been shown that quantum theory is more general than classical theory 
because the classical symmetry algebra can be obtained from the symmetry algebra
in quantum theory by contraction $\hbar\to 0$. 
For these reasons, the most general description in terms of ten-dimensional Lie algebras should be carried out in terms of quantum dS or AdS symmetry. 

The definition of those symmetries is as follows. 
If $M^{ab}$ ($a,b=0,1,2,3,4$, $M^{ab}=-M^{ba}$) are the angular momentum operators for the system under consideration, they should satisfy the
commutation relations:
\begin{equation}
	[M^{ab},M^{cd}]=-i (\eta^{ac}M^{bd}+\eta^{bd}M^{ac}-
	\eta^{ad}M^{bc}-\eta^{bc}M^{ad})
	\label{CR}
\end{equation}
where $\eta^{ab}=0$ if $a\neq b$, 
$\eta^{00}=-\eta^{11}=-\eta^{22}=-\eta^{33}=1$ and 
$\eta^{44}=\mp 1$ for the dS and AdS symmetries, respectively.

Although the dS and AdS groups are the groups of motions of dS and AdS spaces, respectively,
the description in terms of (\ref{CR}) does not involve those groups and spaces,
and {\it it is the definition of dS and AdS symmetries at the quantum level} (see the discussion in \cite{book,DS}). 
In QFT, interacting particles are described by field functions defined on Minkowski, 
dS and AdS spaces. However, as described in Sec. \ref{intro}, the problem statement for PCA  involves only noninteracting bodies
and therefore for PCA we don't need background fields and spaces.

The contraction of the Poincare algebra into the Galilean algebra and the contraction of the quantum algebra into the classical one are widely described in the literature (see, for example, Section 1.3 in \cite{book}). If $c$ is much greater than all velocities in a given system, then Galilean symmetry is a good approximation for describing this system. Similarly, if all angular momenta in a given system are much greater than $\hbar$, then classical physics is a good approximation for describing this system.

In particle theory, the quantities $(c,\hbar)$ are usually not involved and this is characterized such that the system of units $c=\hbar=1$ is used (although the concept of a system of units makes sense only in macroscopic physics). Then all velocities are dimensionless and $\leq 1$ (if tachyons are not taken into account). However, if people want to describe
velocities in $m/s$ then $c$ also has the dimension $m/s$. Physicists
usually understand that physics cannot (and should not) derive that $c\approx 3\cdot 10^8m/s$. This value is purely kinematical (i.e., it does not
depend on gravity and other interactions) and is as is simply because
people want to describe velocities in $m/s$. Since the quantities $(m,s)$ have a physical meaning only at the macroscopic level, one can expect
that the values of $c$ in $m/s$ are different at different stages of the universe. 

Analogously, physicists usually understand that physics cannot (and should not)
derive that $\hbar \approx 1.054\cdot 10^{-34}kg\cdot m^2/s$. This value is purely kinematical
and is as is simply because people want to describe angular momenta
in $kg\cdot m^2/s$. Since the quantities $(kg,m,s)$ have a physical meaning
only at the macroscopic level, one can expect that the values of $\hbar$ in $kg \cdot m^2/s$ are different at different stages of the universe.

Now consider the contraction from dS or AdS symmetry to Poincare one. If the momentum
operators $P^{\nu}$ ($\nu=0,1,2,3$) {\it are defined} as $P^{\nu}=M^{4\nu}/R$ then in the limit when $R\to\infty$, $M^{4\nu}\to\infty$ but the quantities
$P^{\nu}$ are finite, Eqs. (\ref{CR}) become Eqs. (\ref{PCR}). Here {\it R is a parameter which has nothing to do with the dS and AdS spaces}. 
As seen from Eqs. (\ref{CR}), quantum dS and AdS theories
do not involve the dimensional parameters $(c,\hbar,R)$ because $(kg,m,s)$ are meaningful only at the macroscopic level. 

At the quantum level, Eqs. (\ref{CR}) are the most general description
of dS and AdS symmetries and all the operators in Eqs. (\ref{CR}) are dimensionless. At this level,
the theory does not need the quantity $R$ and, by analogy with the choice $(c=\hbar=1)$ in particle theory, $R=1$ is a possible choice. The dimensional quantity $R$ arises if physicists want to deal with the 4-momenta $P^{\mu}$ {\it defined} such that
$M^{4\mu}=RP^{\mu}$. By analogy with the quantities $c$ and $\hbar$, physics 
cannot (and should not) derive the value of $R$. It
is as is simply because people want to measure distances in meters. This value is purely kinematical, i.e., it does not depend on gravity and other interactions. As noted in Sec. \ref{intro},
at the present stage of the universe, $R$ is of the order of $10^{26}m$ but, since the concept of  
meter has a physical meaning only at the macroscopic level, one can expect that the values of $R$ in 
meters are different at different stages of the universe.

Although, at the level of contraction parameters, $R$ has nothing to do with the radius of the background space and is fundamental to the same extent as $c$ and
$\hbar$, physicists usually want to
treat $R$ as the radius of the background space. In GR which is the
non-quantum theory, $\Lambda =\pm 3/R^2$ for
the dS and AdS symmetries, respectively. Physicists usually believe that physics should derive the value 
of $\Lambda$ and that the solution to the dark energy problem depends on this value. 
They also believe that QFT of gravity should confirm the experimental result that, in units
$c=\hbar=1$, $\Lambda$ is of the order
of $10^{-122}/G$ where $G$ is the gravitational constant. This theory can be criticized on several grounds, but the main one is that it is non-renormalizable and contains irremovable divergences.

In any case, the fundamental quantum theory must not contain the quantities 
$(kg,m,s)$ taken from macroscopic theory. 
In particular, quantum theory should not contain the quantities $(c,\hbar,R)$ expressed in terms of $(kg,m,s)$. As noted above, the quantities $(c,\hbar,R)$
can be present in fundamental quantum theory only as contraction parameters for transitions from more general theories to less general ones 
and the question of why these parameters are the way they are and not others does not arise. However,
in QFT, $G$ is not a contraction parameter for transition from a more general theory to a less general one. In \cite{book} we considered a possibility that gravity is a consequence of the fact that finite quantum theory (FQT) based on a finite ring with characteristic $p$ is more general than standard quantum theory (SQT). In that case, 
$G$ depends on $p$ as $1/ln(p)$ and can be considered as a contraction parameter from FQT to SQT when $p\to\infty$.

As noted in Sec. \ref{intro}, in PCA, it is assumed that the bodies are located at large (cosmological) distances from each other and sizes of the bodies are much less than distances between them. Therefore, interactions between the bodies can be neglected and, from the formal point of view, the description of our system
is the same as the description of $N$ free spinless elementary particles.

However, in literature, PCA is usually
considered in the framework of dark energy and other exotic concepts. In Sec. \ref{Dark}
we argue that such considerations are not based on rigorous physical principles. In the present section we have explained how symmetry should be defined at the quantum level and in Sec. \ref{explanation} we briefly sketch how PCA is described the framework of our approach.

\section{Results for cosmological acceleration}
\label{explanation}

As explained above, the most general approach to PCA is to consider this problem within the framework of semiclassical approximation to dS or AdS quantum theory. We first consider the
dS case and the results about the AdS one will be mentioned later. As described in Sec. \ref{intro}, in PCA, the motion of each body can be considered independently of the motion of other bodies. Therefore, the representation  of the dS algebra describing our system is the tensor product of IRs for each body. Since the observed quantities correspond to self-adjoint operators, we must consider selfadjoint IRs of the dS algebra.

Unitary IRs of the dS group have been considered by several authors. By using the results of the
excellent Mensky's book \cite{Mensky}, we described selfadjoint IRs of the dS algebra in \cite{JPA,JMP,symm}. We will consider the operators $M^{4\mu}$ not
only in Poincare approximation but also taking into account dS
corrections. If those corrections are small, then, as explained in \cite{Axioms},
IRs under consideration can be described by Eqs. (2.2) in that reference. 

These equations describe IRs in momentum representation and {\it at this stage, we have no spatial coordinates yet}. However, in the semiclassical approximation it is necessary to know how the momentum representation is related to the coordinate one. 
These representations are usually considered to be related by the Fourier transform.
As shown in \cite{book}, such a connection is not universal, for example it does not work for photons from distant stars. However, since bodies in PCA can be described in the nonrelativistic approximation, the position operator in momentum representation
can be defined as usual, i.e., as ${\bf r}=i\hbar \partial /\partial {\bf p}$. 

In semiclassical approximation, we can treat ${\bf p}$ and ${\bf r}$ as usual vectors. 
Then as follows from Eqs. (2.2) in \cite{Axioms}
\begin{equation}
{\bf P}= {\bf p}+mc{\bf r}/R, \quad H = {\bf p}^2/2m +c{\bf p}{\bf r}/R,\quad {\bf N}=-m{\bf r}
\label{II64}
\end{equation}
where $H=E-mc^2$ is the classical nonrelativistic Hamiltonian and ${\bf N}=(M^{01},M^{02},M^{03})$
is the operator of Lorentz boosts.
As follows from these expressions and Eqs. (2.2) in \cite{Axioms}
\begin{equation}
H({\bf P},{\bf r})=\frac{{\bf P}^2}{2m}-\frac{mc^2{\bf r}^2}{2R^2}
\label{II66}
\end{equation}
where the last term is the dS correction to the non-relativistic Hamiltonian. 
A shown in \cite{Axioms}, now it follows from the Hamilton equations 
that a free particle is moving with the acceleration
\begin{equation}
{\bf a}={\bf r}c^2/R^2=\frac{1}{3}c^2\Lambda {\bf r}
\label{accel}
\end{equation}
where ${\bf r}$ is the
radius vector of the particle and $\Lambda=3/R^2$. 

To describe a system of $N$ bodies, it is necessary to take into account that it is described by the tensor product of single-body representations. Therefore, each operator $M^{ab}$ for the $N$-body system is the sum of the corresponding single-body operators $M^{ab}$. Then, as shown in \cite{Axioms},  the relative
acceleration also is given by Eq. (\ref{accel}) but now
${\bf a}$ is the relative acceleration and ${\bf r}$ is the
relative radius vector, i.e., Eq. (\ref{Perl}) is indeed valid. 

As noted in \cite{Axioms}, dS symmetry is more general than AdS one.
Formally, an analogous calculation using the results of Chap. 8 of
\cite{book} on IRs of the AdS algebra gives
that, in the AdS case, ${\bf a}=-{\bf r}c^2/R^2$, i.e., we have attraction instead of repulsion.  The experimental facts that the bodies repel each other confirm that dS symmetry is indeed more
general than AdS one. 

The relative accelerations given by (\ref{Perl}) are formally the same as those derived from GR if the curvature of dS space equals $\Lambda=3/R^2$, where $R$ is the radius of this space. 
{\it However, the crucial difference between our results and the results of GR is as follows.
While in GR, $R$ is the radius of the dS space and can be arbitrary, as explained in detail in Sec.
\ref{Hierarchy}, in quantum
theory, $R$ has nothing to do with the
radius of the dS space, it is the coefficient of proportionality between $M^{4\mu}$ and $P^{\mu}$, it is fundamental to the same extent as $c$ and $\hbar$, and a question why $R$ is as is does not arise.} {\bf Therefore, our approach gives a clear explanation why $\Lambda$ is
as is.}

In literature, it is often stated that quantum theory of gravity should become GR in classical approximation.
In Sec. \ref{Hierarchy} we argue that this is probably not the case because at the quantum level
the concept of space-time background does not have a physical meaning. {\it Our results for the cosmological acceleration obtained
from semiclassical approximation to quantum theory are compatible with GR} but in our approach, space-time background is absent from the very beginning.

{\bf Conclusion}: {\it The problem of cosmological acceleration has a unique solution which has nothing to do with dark energy or other artificial reasons: cosmological acceleration is an inevitable {\bf kinematical} consequence of quantum theory in semiclassical approximation.}

{\bf Acknowledgments}. I am grateful to Vladimir Karmanov, Dmitri Kazakov, Michel Planat and Ethan Vishniac for important useful comments.

\end{document}